\documentclass[twocolumn,aps,prd,showpacs]{revtex4}
\usepackage{graphics,graphicx}
\usepackage{amssymb,amsfonts}

\def\noi{\noindent}

\def\nq{\hspace*{-1em}}

\def\nhh{\hspace*{-0.3em}}
\def\cm{\hspace*{1cm}}

\def\Jl#1#2{{\it #1\/} {\bf #2},\ }

\def\ApJ#1 {\Jl{Astroph. J.}{#1}}
\def\CQG#1 {\Jl{Class. Quantum Grav.}{#1}}
\def\DAN#1 {\Jl{Dokl. AN SSSR}{#1}}
\def\GC#1 {\Jl{Grav. \& Cosmol.}{#1}}
\def\GRG#1 {\Jl{Gen. Rel. Grav.}{#1}}
\def\JETF#1 {\Jl{Zh. Eksp. Teor. Fiz.}{#1}}
\def\JETP#1 {\Jl{Sov. Phys. JETP}{#1}}
\def\JHEP#1 {\Jl{JHEP}{#1}}
\def\JMP#1 {\Jl{J. Math. Phys.}{#1}}
\def\NPB#1 {\Jl{Nucl. Phys.}{B\ #1}}
\def\NP#1 {\Jl{Nucl. Phys.}{#1}}
\def\PLA#1 {\Jl{Phys. Lett.}{#1A}}
\def\PLB#1 {\Jl{Phys. Lett.}{#1B}}
\def\PRD#1 {\Jl{Phys. Rev.}{D\ #1}}
\def\PRL#1 {\Jl{Phys. Rev. Lett.}{#1}}

\def\lal{&& {}\nhh}
\def\eq{Eq.\,}
\def\eqs{Eqs.\,}
\def\beq{\begin{equation}}
\def\eeq{\end{equation}}
\def\bear{\begin{eqnarray}}
\def\bearr{\begin{eqnarray} \lal}
\def\ear{\end{eqnarray}}
\def\earn{\nonumber \end{eqnarray}}

\def\nnn{\nonumber\\ \lal }

\def\eql{&\! = &\!}

\def\d{\partial}

\def\const{{\rm const}}
\def\eps{\varepsilon}

\def\N{{\mathbb N}}
\def\R{{\mathbb R}}

\def\MN{^{\mu\nu}}

\def\wh{wormhole}
\def\whs{wormholes}
\def\bh{black hole}
\def\bhs{black holes}

\def\ssph{static, spherically symmetric}
\def\asflat{asymptotically flat}
\def\asdS{asymptotically de Sitter}

\def\Sch{Schwarz\-schild}

\begin{document}

\title{Regular phantom black holes}

\author{K.A. Bronnikov}
\affiliation
    {Center for Gravitation and Fundamental Metrology, VNIIMS, 46 Ozyornaya St., Moscow, Russia;
     Institute of Gravitation and Cosmology, PFUR, 6 Miklukho-Maklaya St., Moscow 117198, Russia.
         E-mail: kb20@yandex.ru}

\author{J.C. Fabris}
\affiliation
   {Departamento de F\'{\i}sica, Universidade Federal do Esp\'{\i}rito Santo,
      Vit\'oria, 29060-900, Esp\'{\i}rito Santo, Brazil.
      E-mail: fabris@cce.ufes.br}

\begin{abstract}
    For self-gravitating, \ssph, minimally coupled scalar fields with
    arbitrary potentials and negative kinetic energy (favored by the
    cosmological observations), we give a classification of possible regular
    solutions to the field equations with flat, de Sitter and AdS asymptotic
    behavior. Among the 16 presented classes of regular rsolutions are
    traversable wormholes, Kantowski-Sachs (KS) cosmologies beginning and
    ending with de Sitter stages, and \asflat\ \bhs\ (BHs). The Penrose
    diagram of a regular BH is Schwarzschild-like, but the singularity at
    $r=0$ is replaced by a de Sitter infinity, which gives a hypothetic BH
    explorer a chance to survive. Such solutions also lead to the idea that
    our Universe could be created from a phantom-dominated collapse in
    another universe, with KS expansion and isotropization after crossing
    the horizon. Explicit examples of regular solutions are built and
    discussed. Possible generalizations include $k$-essence type scalar
    fields (with a potential) and scalar-tensor theories of gravity.


\end{abstract}
\maketitle


    Observations provide more and more evidence that the modern accelerated
    expansion of our Universe is governed by a peculiar kind of matter,
    called dark energy (DE), characterized by negative values of the
    pressure to density ratio $w$. Moreover, by current estimates, even
    $w < -1$ seems rather likely
    \cite{steinhardt,tegmark,seljak,hannestad,star03,chandra},
    though many of such estimates are model-dependent. Thus, assuming a
    perfect-fluid DE with $w = \const$ implies, using various observational
    data (CMB, type Ia supernovae, large-scale structure), $- 1.39 < w <
    -0.79$ at $2\sigma$ level \cite{hannestad}. Considerations of a variable
    DE equation of state \cite{steinhardt,tegmark} also allow highly
    negative values of $w$. A model-independent study \cite{star03} of data
    sets containing 172 SNIa showed a preferable range $-1.2 < w < -1$ for
    the recent epoch. Similar figures follow from an analysis of the Chandra
    telescope observations of hot gas in 26 X-ray luminous dynamically
    relaxed galaxy clusters \cite{chandra}: $w =-1.20^{+0.24}_{-0.28}$.

    Moreover, a highly negative $w$ makes negligible the undesirable DE
    contribution to the total energy density in the period of structure
    formation. Thus, even if the cosmological constant, giving precisely $w
    = -1$, is still admitted by observations as possible DE, there is a need
    for a more general framework allowing $w < -1$.

    The perfect-fluid description of DE is plagued with instability at small
    scales due to imaginary velocity of sound; more consistent descriptions
    providing $w < -1$ use self-interacting scalar fields with negative
    kinetic energy (phantom scalars) or tachyonic fields
    \cite{sen,gorini,fara05} (see also references therein). To avoid the
    obvious quantum instability, a phantom scalar may perhaps be regarded as
    an effective field description following from an underlying theory with
    positive energies \cite{no03,trod}. Curiously, in a classical setting, a
    massless phantom field even shows a more stable behavior than its usual
    couterpart \cite{cold,picon}. A fundamental origin of phantom fields is
    under discussion, but they naturally appear in some models of string
    theory \cite{sen}, supergravities \cite{sugr} and theories in more than
    11 dimensions like F-theory \cite{khvie}.

    If a phantom scalar, be it basic or effective, is part of the real
    field content of our Universe, it is natural to seek its manifestations
    not only in cosmology but also in local phenomena, in particular, in
    \bh\ (BH) physics, as, e.g., in the recent works on DE accretion onto
    BHs \cite{ero,a_frol} and on BH interaction with a phantom shell
    \cite{berez05}.

    We are trying here to find out which kinds of regular \ssph\
    configurations may be formed by a phantom scalar field itself. Since it
    does not respect the usual energy conditions, nonsingular solutions of
    particular physical interest for BH physics and/or cosmology could be
    expected. Our main finding is, in our view, the existence of regular
    \asflat\ BH solutions with an expanding, asymptotically de Sitter
    Kantowski-Sachs (KS) cosmology beyond the event horizon. It is, to our
    knowledge, quite a new way of avoiding a BH central singularity,
    alternative to known solutions with a regular center (see, e.g.,
    \cite{dym92,ned01,bdd03}).

    The plan is as follows. After writing the field equations, we mention
    some no-go theorems for both normal and phantom scalars (without
    proofs), making sure that they leave sufficient freedom for solutions of
    interest. Then follows a simple qualitative analysis which reveals 16
    classes of possible nonsingular solutions and the properties of the
    potential necessary for their existence. Using the inverse problem
    method, we construct simple explicit examples of different types of
    solutions, which include, among others, regular KS cosmologies, \asdS\
    both in the past and in the future, and regular \bhs, \asflat\ or
    anti-de Sitter (AdS) in the static (R) region and \asdS\ in the
    nonstatic (T) region. Some generalizations of these results are
    indicated in conclusion.

    We start with the action for a self-gravitating minimally coupled scalar
    field with an arbitrary potential $V(\phi)$
\beq                                                            \label{act}
     S = \int \sqrt{g}\, d^4 x [R + \eps g\MN\d_\mu\phi \d_\nu\phi
     			- 2V(\phi)],
\eeq
    where $R$ is the scalar curvature, $\eps=+1$ corresponds to a usual
    scalar field with positive kinetic energy and $\eps=-1$ to a phantom
    field. For the general \ssph\ metric
\beq                                                            \label{ds}
    ds^2 = A(\rho) dt^2 - \frac{d\rho^2}{A(\rho)} -
    		r^2(\rho) (d\theta^2 + \sin^2\theta\, d\varphi^2),
\eeq
    and $\phi=\phi(\rho)$, the scalar field equation and three independent
    combinations of the Einstein equations read
\bear
	 (Ar^2 \phi')' \eql \eps r^2 dV/d\phi,                  \label{phi}
\\
              (A'r^2)' \eql - 2r^2 V;                           \label{00}
\\
              2 r''/r \eql -\eps{\phi'}^2 ;                     \label{01}
\\
         A (r^2)'' - r^2 A'' \eql 2 ,                           \label{02}
\ear
    where the prime denotes $d/d\rho$. \eq (\ref{phi}) follows from
    (\ref{00})--(\ref{02}), which, given a potential $V(\phi)$, form a
    determined set of equations for the unknowns $r(\rho),\ A(\rho),\
    \phi(\rho)$. \eq(\ref{02}) can be once integrated giving
\bear
    B' \equiv \biggl(\frac{A}{r^2}\biggr)'
    			= \frac{2(\rho_0 -\rho)}{r^4},          \label{B'}
\ear
    where $B(\rho) = A/r^2$ and $\rho_0$ is an integration constant.

    The coordinate $\rho$ has been chosen in such a way that Killing
    horizons, if any, correspond to regular zeros of the function $A(\rho)$:
    $A(\rho) \approx (\rho-\rho_h)^p$ where $p\in \N$ is the order of the
    horizon \cite{cold}. The metric is static where $A(\rho) > 0$ (in R
    regions), while where $A < 0$ (in T regions) $\rho$ is a time
    coordinate, and (\ref{ds}) describes a homogeneous anisotropic KS
    cosmology.

    Some general consequences of \eqs (\ref{phi})--(\ref{02}) (no-go
    theorems) constrain the nature of possible solutions.

    (A). For $\eps=+1$ one cannot obtain wormholes or configurations ending
    with a regular 3-cylinder of finite radius $r$ \cite{vac1}. This result
    follows from \eq (\ref{01}) (giving $r'' \leq 0$) and is valid
    independently of the large $r$ behaviour of the metric --- flat, AdS or
    any other. For phantom fields \eq (\ref{01}) gives $r'' \geq 0$, and
    such a restriction is absent. Thus, for a free massless phantom field
    wormhole solutions are well known since the 70s \cite{br73,h_ell}.

    (B). Particlelike (or starlike) solutions (PLS), i.e., \asflat\
    solutions with a regular center, are not excluded for both kinds of
    scalar fields but under certain constraints on the potential. Thus,
    {\it for $\eps=+1$, PLS cannot be obtained with $V(\phi)\geq 0$}
    \cite{vac5}.

    For $\eps = -1$, on the contrary, no PLS can appear if $V(\phi) \leq 0$,
    and PLS with mass $m > 0$ require a potential of alternating sign.
    The (Schwarzschild) mass $m$ is defined here by assuming $r\approx \rho$
    and $A \approx 1-2m/\rho$ as $\rho\to\infty$.

    (C). A no-hair theorem, extending the one known for $\eps=+1$
    \cite{bek,ad-pear}, sounds as follows: {\it The only \asflat\ BH
    solution to \eqs (\ref{00})--(\ref{02}) is characterized by $\phi =
    \const$ and the \Sch\ metric in the whole domain of outer communication,
    if either (i) $\eps=+1$ and $V(\phi)\geq 0$ or (ii) $\eps = -1$,
    $V(\phi)\leq 0$, and $r' > 0$ at and outside the event horizon.}

    It will be seen below that there exist BH solutions where the condition
    $r' > 0$ is violated.

    (D). \eq (\ref{B'}) severely restricts the possible dispositions of
    Killing horizons in the resulting metric and consequently the global
    causal structure of space-time \cite{vac1}.

    Indeed, horizons are regular zeros of $A(\rho)$ and hence $B(\rho)$.
    By (\ref{B'}), $B(\rho)$ increases at $\rho < \rho_0$, has a maximum at
    $\rho=\rho_0$ and decreases at $\rho>\rho_0$. It can have {\it at
    most\/} two simple zeros, bounding a range $B > 0$ (R region), or one
    double zero and two T regions around. It can certainly have a single
    simple zero or no zeros at all.

    So the choice of possible types of global causal structure is precisely
    the same as for the general Schwarzschild-de Sitter solution with
    arbitrary mass and cosmological constant.

    \eq (\ref{02}) does not contain $\eps$, hence this result ({\it the
    Global Structure Theorem\/} \cite{vac1}) equally applies to normal and
    phantom fields. It holds for any sign and shape of $V(\phi)$ and
    under any assumption on the asymptotics. In particular, BHs with scalar
    hair (certainly, respecting the no-hair theorems) are not excluded. Some
    examples of (singular) BHs with both normal (e.g., \cite{vac2,mann95})
    and phantom \cite{lecht95,lecht96} scalar hair are known. However,
    \asflat\ BHs with a regular center are ruled out.

    The Hawking temperature of a horizon $\rho = h$ is determined
    \cite{wald} as $T_{\rm H} = \kappa/(2\pi)$ where $\kappa$ is the surface
    gravity at $\rho = h$. In our system,
\beq                                                           \label{hawk}
      \kappa =|A'(h)|/2\quad {\rm and} \quad A'(h) = (\rho_0-h)/r^2 (h).
\eeq

    Let us indicate the possible kinds of nonsingular solutions without
    restricting the shape of $V(\phi)$. Assuming no pathology at
    intermediate $\rho$, regularity is determined by the system behavior at
    the ends of the $\rho$ range. The latter may be classified as a regular
    infinity ($r \to \infty$), which may be flat, de Sitter or AdS (other
    variants, like $r^2\sim \rho$, can exist but seem to be of lesser
    interest), a regular center, and the intermediate case $r \to r_0 > 0$.
    Any kind of oscillatory behavior of $r(\rho)$ is ruled out by the
    constant sign of $r''$.

    Suppose we have a regular infinity as $\rho \to \infty$, so that
    $V\to V_+ =\const$ while the metric becomes Minkowski (M), de Sitter
    (dS) or AdS according to the sign of $V_+$. In all cases $r \approx
    \rho$ at large $\rho$.

    For $\eps=+1$, due to $r'' \leq 0$, $r$ necessarily vanishes at some
    $\rho=\rho_c$, which means a center, and the only possible regular
    solutions interpolate between a regular center and an AdS, flat or dS
    asymptotic; in the latter case the causal structure coincides with that
    of de Sitter space-time.

    For $\eps=-1$, there are similar solutions with a regular center,
    but due to $r''\geq 0$ one may assume $\rho \in \R$ and obtain either
    $r\to r_0 =\const > 0$ or $r \to\infty$ as $\rho\to -\infty$. In other
    words, all kinds of regular behavior are possible at the other end.
    In particular, if $r\to r_0$, we get $A \approx -\rho^2/r_0^2$, i.e., a
    T region comprising a highly anisotropic KS cosmology with one scale
    factor ($r$) tending to a constant while the other ($A$) inflates. The
    scalar field tends to a constant, while $V(\phi) \to 1/r_0^2$.

    Thus there are three kinds of regular asymptotics at one end,
    $\rho\to\infty$ (M, dS, AdS), and four at the other, $\rho\to -\infty$:
    the same three plus $r\to r_0$, simply $r_0$ for short. (The asymmetry
    has appeared since we did not allow $r\to\const$ as $\rho\to\infty$.
    The inequality $r''> 0$ forbids nontrivial solutions with two such
    $r_0$-asymptotics.) This makes nine combinations shown in Table 1.
    Moreover, each of the two cases labelled KS* actually comprises
    three types of solutions according to the properties of $A(\rho)$: there
    can be two simple horizons, one double horizon or no horizons between
    two dS asymptotics. Recalling 3 kinds of solutions with a regular
    center, we obtain as many as 16 qualitatively different classes of
    globally regular configurations of phantom scalar fields.

\begin{table}[h]
    \caption{Regular solutions with $\rho\in\R$ for $\eps=-1$. Each row
    corresponds to a certain asymptotic behavior as $\rho\to +\infty$, each
    column --- to $\rho\to -\infty$. The mark ``sym'' refers to combinations
    obtained from others by symmetry $\rho \leftrightarrow -\rho$.}
\begin{center}
\begin{tabular}{|c|c|c|c|c|}
\hline
	      &	  AdS    &    M     &   dS   &  $r_0$    \\
\hline
       AdS    &  \wh     &   \wh    &   \bh  &  \bh      \\
\hline
       M      &  sym     &   \wh    &   \bh  &  \bh      \\
\hline
       dS     &  sym     &   sym    &   KS*  &   KS*     \\
\hline
\end{tabular}
\end{center}
\end{table}

    Examples of each behavior may be found in an algorithmic manner by
    properly choosing the function $r(\rho)$ and invoking the inverse
    problem method:  $B(\rho)$ and $A(\rho)$ are then obtained from \eq
    (\ref{B'}) (and $B(\rho)$ always behaves as described above), after that
    $\phi(\rho)$ is yielded by \eq (\ref{01}) and $V(\rho)$ by \eq
    (\ref{00}). A critical requirement is that $r(\rho)$ must satisfy the
    inequality $r''\leq 0$ for $\eps=1$ and $r''\geq 0$ for $\eps=-1$. The
    function $V(\phi)$ is restored from known $V(\rho)$ and $\phi(\rho)$
    provided the latter is monotonic, which is the case if everywhere
    $r''\ne 0$.

    The potential $V$ tends to a constant and, moreover, $dV/d\phi \to 0$ at
    each end of the $\rho$ range. Therefore any model from the above classes
    requires a potential with at least two zero-slope points (not
    necessarily extrema) at different values of $\phi$. Suitable potentials
    are, e.g., $V = V_0 \cos^2 (\phi/\phi_0)$ and the Mexican hat potential
    $V = (\lambda/4)(\phi^2 - \eta^2)^2$ where $V_0,\ \phi_0,\ \lambda,\
    \eta$ are constants. A flat infinity certainly requires $V_+ = 0$, while
    a de Sitter asymptotic can correspond to a maximum of $V$ since phantom
    fields tend to climbing up the slope of the potential rather than
    rolling down, as is evident from \eq (\ref{phi}). Accordingly,
    Faraoni \cite{fara05}, considering spatially flat isotropic phantom
    cosmologies, has shown that if $V(\phi)$ is bounded above by
    $V_0 = \const > 0$, the de Sitter solution is a global attractor. Very
    probably this conclusion extends to KS cosmologies after isotropization.

    We will now give a transparent analytic example, leaving for the future
    more elaborated models with better motivated potentials.
    So we put $\eps=-1$,
\beq                                                         \label{r1}
	r = (\rho^2 + b^2)^{1/2}, \cm b = \const > 0.
\eeq
    and use the inverse problem scheme. \eq (\ref{B'}) gives
\bearr  \nq                                                   \label{B1}
      B(\rho) = A(\rho)/r^2(\rho)
\nnn
      = \frac{c}{b^2} + \frac{1}{b^2+\rho^2} + \frac{\rho_0}{b^3}
	\left(\frac{b\rho}{b^2 + \rho^2} + \tan^{-1}\frac{\rho}{b}\right),
\ear
    where $c = \const$. \eqs (\ref{01}) and (\ref{00}) then lead to
    expressions for $\phi(\rho)$ and $V(\rho)$:
\bear                                                        \label{phi1}
      \phi \eql \pm\sqrt{2} \tan^{-1} (\rho/b) + \phi_0,
\\                                                           \label{V1}
      V \eql - \frac{c}{b^2} \left(\!1 + \frac{2\rho^2}{r^2}\right)
	   - \frac{\rho_0}{b^3}
	   	\left[ \frac{3b\rho}{r^2}
    + \left(1 + \frac{2\rho^2}{r^2}\right)\tan^{-1}\frac{\rho}{b}\right]
\nnn
\ear
    with $r=r(\rho)$ given by (\ref{r1}). In particular,
\beq                                                       \label{BV_as}
      B(\pm \infty) = -\frac 13 V(\pm \infty)
			= \frac{2bc \pm \pi \rho_0}{2b^3}.
\eeq
    Choosing in (\ref{phi1}), without loss of generality, the plus sign and
    $\phi_0=0$, we obtain for $V(\phi)$ ($\psi := \phi/\sqrt{2}$):
\bearr   \nq
     V(\phi) = -\frac{c}{b^2} (3 - 2\cos^2 \psi)             \label{Vf}
\nnn \ \
     - \frac{\rho_0}{b^3} \left[3\sin\psi \cos\psi
         			+ \psi (3 - 2\cos^2 \psi) \right].
\ear

    The solution behavior is controlled by two integration constants: $c$
    that moves $B(\rho)$ up and down, and $\rho_0$ showing the maximum of
    $B(\rho)$. Both $r(\rho)$ and $B(\rho)$ are even functions if
    $\rho_0=0$, otherwise $B(\rho)$ loses this symmetry.

    In the simplest case $\rho_0 = c =0$ we obtain the so-called Ellis \wh\
    \cite{h_ell}: $V \equiv 0$ and $A \equiv 1$.

    Solutions with $\rho_0 = 0$ but $c\ne 0$ describe symmetric structures:
    \whs\ with two AdS asymptotics if $c > 0$ and solutions with two dS
    asymptotics if $c <0$. If $0> c > -1$, there is an R region in the
    middle, bounded by two simple horizons, at $c=-1$ they merge into a
    double horizon, and $c < -1$ leads to a pure KS cosmology.

    With $\rho_0 \ne 0$, the two asymptotics look differently. Let us
    dwell upon solutions which are flat at $\rho = \infty$. Then the
    constants obey the condition $2bc = -\pi \rho_0$, while the Schwarzschild
    mass is $m =\rho_0/3$. According to (\ref{BV_as}), for $\rho_0 <0$ ($c >
    0$) we obtain a \wh\ with $m < 0$ and an AdS metric at the far end,
    corresponding to the cosmological constant $V_- < 0$. For $\rho_0 >0$,
    when $V_{-} > 0$, there is a regular BH with $m > 0$ and a dS asymptotic
    far beyond the horizon. As any \asflat\ BH with a simple horizon, it has
    a Schwarzschild-like causal structure, but the singularity $r=0$ in the
    Carter-Penrose diagram is replaced by $r=\infty$.

    The horizon radius depends on both parameters $m$ and $b = \min r(\rho)$
    and cannot be smaller than $b$, which also plays the role of a scalar
    charge:  $\psi \approx \pi/2 -b/\rho$ at large $\rho$. Since $A(0) =
    1+c$, the throat $\rho=0$ is located in the R region if $c >-1$, i.e.,
    if $3\pi m < 2b$, at the horizon if $3\pi m = 2b$ and in the T region
    beyond it if $3\pi m > 2b$.

    Such regular BHs combine the properties of BHs, whose main feature is a
    horizon, and \whs, whose main feature is a throat, $r= r_{\min} >0$. The
    above relations between $m$ and $b$ show (and it is probably generically
    true) that if the BH mass dominates over the scalar charge, the throat
    is invisible to a distant observer, and the BH looks almost as usual in
    general relativity. However, a possible BH explorer now gets a chance to
    survive for a new life in an expanding KS universe.

    One may also speculate that our Universe could appear from collapse to a
    phantom BH in another, ``mother'' universe and undergo isotropization
    (e.g., due to particle creation) soon after crossing the horizon. The KS
    nature of our Universe is not excluded observationally \cite{craw} if
    its isotropization had happened early enough, before the last scattering
    epoch (at redshifts $z\gtrsim 1000$). The same idea of a Null Bang
    instead of a Big Bang (cosmological expansion starting from a horizon
    rather than a singularity) was discussed in \cite{bdd03} for a system
    with a de Sitter vacuum core and a regular center in the R region.

    Let us note in conclusion that the present analysis, which has revealed
    a wealth of regular solutions including BHs, is easily extended to more
    sophisticated phantom models, e.g., to those of $k$-essence type.
    Indeed, for the scalar field Lagrangian $L= P(X) - 2V(\phi)$ where $X =
    g\MN \phi_{,\mu} \phi_{,\nu}$ and $P$ is an arbitrary function, \eqs
    (\ref{02}) and (\ref{B'}) remain unchanged while the crucial inequality
    $r'' \geq 0$ holds if the theory satisfies the ``phantom condition''
    $dP/dX < 0$. $k$-essence type theories, among other merits, are known to
    avoid inadmissible sound velocities and the stabilization problem
    \cite{scherrer,lazcoz}.

    Other obvious generalizations are scalar-tensor theories of gravity
    and, as their subclass, nonminimally coupled scalar fields with
    Lagrangians including $\const \cdot R \phi^2$. Such theories are
    conformally related to (\ref{act}), and the conformal factors, if
    well-behaved, do not change the causal and asymptotic properties of the
    solutions.

\medskip\noi
    {\bf Acknowledgments.}
    We are thank Nelson Pinto-Neto and J\'er\^ome Martin for helpful
    discussions. KB thanks the colleagues from DF-UFES for hospitality. The
    work was supported by CNPq (Brazil); KB was also supported by ISTC
    Project 1655.

\end{document}